\documentclass[showpacs,aps,pra,amsmath,amssymb,floatfix]{revtex4}
\preprint{draft 1.5}

\usepackage{graphicx}
\usepackage{dcolumn}
\usepackage{bm}
\usepackage{amssymb}
\usepackage{setspace}

\usepackage{epsfig}
\usepackage[T1]{fontenc}

\allowdisplaybreaks[1]

\begin{document}

\title{Relevance of various Dirac covariants in hadronic Bethe-Salpeter 
wave functions in electromagnetic decays of ground state vector mesons}

\author{Shashank Bhatnagar$^1$}
\affiliation{Department of Physics, Addis Ababa University,\\
P.O. Box 1148/1110, Addis Ababa, Ethiopia}
\author{Jorge Mahecha$^2$}
\affiliation{Instituto de F\'{\i}sica, Universidad de Antioquia UdeA;\\
Calle 70 No. 52-21, Medell\'{\i}n, Colombia}
\author{Yikdem Mengesha}
\affiliation{Department of Physics, Addis Ababa University,\\
P.O. Box 1148/1110, Addis Ababa, Ethiopia}


\date{\today}


\begin{abstract}

In this work we have employed Bethe-Salpeter equation (BSE) under 
covariant instantaneous ansatz (CIA) to study electromagnetic decays of 
ground state equal mass vector mesons: $\rho$, $\omega$, $\phi$, $\psi$ 
and $Y$ through the process $V\rightarrow\gamma*\rightarrow e^+ + e^-$. 
We employ the generalized structure of hadron-quark vertex function 
$\Gamma$ which incorporates various Dirac structures from their complete 
set order-by-order in powers of inverse of meson mass. The 
electromagnetic decay constants for the above mesons are calculated 
using the leading order (LO) and the next-to-leading order (NLO) Dirac 
structures. The relevance of various Dirac structures in this 
calculation is studied.

\end{abstract}



\maketitle

\noindent\footnoterule

\noindent{\footnotesize $^1$ Corresponding author. \tt shashank$_-$bhatnagar@yahoo.com}\\
\noindent{\footnotesize $^2$ \tt mahecha@fisica.udea.edu.co}


\section{Introduction}
\label{sec:intro}
Meson decays provide an important tool for exploring the structures of 
these simplest bound states in QCD, and for studies on non-perturbative 
behavior of strong interactions. These studies has become a hot topic in 
recent years. Flavourless vector mesons play an important role in hadron 
physics due to their direct coupling to photons and thus provide an 
invaluable insight into the phenomenology of electromagnetic couplings 
to hadrons. Thus, a realistic description of vector mesons at the quark 
level of compositeness would be an important element in our 
understanding of hadron dynamics and reaction processes at scales where 
QCD degrees of freedom are relevant. There have been a number of 
studies\cite{alkofer02,ivanov99,close02,li08,cvetic04,hwang97,alkofer01,wang08} 
on processes involving strong, radiative and leptonic decays of vector 
mesons. Such studies offer a direct probe of hadron structure and help 
in revealing some aspects of the underlying quark-gluon dynamics.

In this work we study electromagnetic decays of ground state equal mass 
vector mesons: $\rho,\omega,\phi,\psi$ and $Y$ (each comprising of equal 
mass quarks) through the process $V\rightarrow\gamma*\rightarrow e + e-$ 
which proceeds through the coupling of quark-anti quark loop to the 
electromagnetic current in the framework of Bethe-Salpeter Equation 
(BSE), which is a conventional non-perturbative approach in dealing with 
relativistic bound state problems in QCD and is firmly established in 
the framework of Field Theory. From the solutions, we obtain useful 
information about the inner structure of hadrons which is also crucial 
in high energy hadronic scattering and production processes. Despite its 
drawback of having to input model dependent kernel, these studies have 
become an interesting topic in recent years since calculations have 
satisfactory results as more and more data is being accumulated. We get 
useful insight about the treatment of various processes using BSE due to 
the unambiguous definition of the 4D BS wave function which provides 
exact effective coupling vertex (Hadron-quark vertex) of the hadron with 
all its constituents (quarks).

We have employed QCD motivated Bethe-Salpeter Equation (BSE) under 
Covariant Instantaneous Ansatz 
(CIA)\cite{mitra01,mitra99,bhatnagar91,mitra92,bhatnagar06,bhatnagar11} 
to calculate this process. CIA is a Lorentz-invariant generalization of 
Instantaneous Ansatz (IA). What distinguishes CIA from other 3D 
reductions of BSE is its capacity for a two-way interconnection: an 
exact 3D BSE reduction for a $q\overline{q}$ system (for calculation of 
mass spectrum), and an equally exact reconstruction of original 4D BSE 
(for calculation of transition amplitudes as 4D quark loop integrals). 
In these studies, the main ingredient is the 4D hadron-quark vertex 
function $\Gamma$ which plays the role of an exact effective coupling 
vertex of the hadron with all its constituents (quarks). The complete 4D 
BS wave function $\Psi(P,q)$ for a hadron of momentum $P$ and internal 
momentum $q$ comprises of the two quark propagators (corresponding to 
two constituent quarks) bounding the hadron-quark vertex $\Gamma$. This 
4D BS wave function is considered to sum up all the non-perturbative QCD 
effects in the hadron. Now one of the main ingredients in 4D BS wave 
function (BSW) is its Dirac structure. The copius Dirac structure of BSW 
was already studied by C.H.L. Smith\cite{smith69} much earlier. Recent 
studies\cite{alkofer02,li08,cvetic04} have revealed that various mesons 
have many different Dirac structures in their BS wave functions, whose 
inclusion is necessary to obtain quantitatively accurate observables. It 
was further noticed that all structures do not contribute equally for 
calculation of various meson observables\cite{alkofer02,alkofer01}. 
Further, it was amply noted in \cite{hawes98} that many hadronic 
processes are particularly sensitive to higher order Dirac structures in 
BS amplitudes. It was further noted in \cite{hawes98} that inclusion of 
higher order Dirac structures is also important to obtain simultaneous 
agreement with experimental decay widths for a range of processes such 
as: $V\rightarrow e^{+}e^{-}$, $V\rightarrow \gamma P$, $V\rightarrow 
PP$, etc. for a given choice of parameters.

Towards this end, to ensure a systematic procedure of incorporating 
various Dirac covariants from their complete set in the BSWs of various 
hadrons (pseudoscalar, vector etc.), we developed a naive power counting 
rule in ref.\cite{bhatnagar06}, by which we incorporate various Dirac 
structures in BSW, order-by-order in powers of inverse of meson mass. 
Using this power counting rule we calculated electromagnetic decay 
constants of vector mesons ($\rho,\omega,\phi$) using only the leading 
order (LO) Dirac structures [$i\gamma\cdot\varepsilon$ and 
$(\gamma\cdot\varepsilon)(\gamma\cdot P)/M$]. However in 
Ref.\cite{bhatnagar11}, we rigorously studied leptonic decays of unequal 
mass pseudoscalar mesons $\pi,K,D,D_{s},B$ and calculated the leptonic 
decay constants $f_{P}$ for these mesons employing both the leading 
order (LO) and the next-to-leading order (NLO) Dirac structures. The 
contributions of both LO and NLO Dirac structures to $f_{P}$ was worked 
out. We further studied the relevance of both the LO and the NLO Dirac 
structures to this calculation. In the present paper, we extended these 
studies to vector mesons and have employed both LO and NLO Dirac 
structures identified according to our power counting rule, to calculate 
$f_{V}$ for ground state vector mesons, $\rho,\omega,\phi,\psi,Y$ and in 
the process we studied the relevance of various Dirac structures to 
calculation of decay constants $f_{V}$ for vector mesons in the process 
$V\rightarrow e^{+}e^{-}$. We found that contributions from NLO Dirac 
structures are smaller than those of LO Dirac structures for all vector 
mesons. In what follows, we give a detailed discussion of the fit and 
calculation up to NLO level after a brief review of our framework.

The paper is organized as follows: In section 2 we discuss the structure 
of BSW for vector mesons under CIA using the power counting rule we 
proposed earlier. In section 3 we give the calculation of $f_{V}$ for 
vector mesons. A detailed presentation of results and the numerical 
calculation is given in section 4. Section 5 is relegated to Discussion.


\section{BSE under CIA} \label{sec:theory} We briefly outline the BSE
framework under CIA. For simplicity, lets consider a
$\mathrm{q}\overline{\mathrm{q}}$ system comprising of scalar quarks
with an effective kernel $K$, 4D wave function $\Phi(P,q)$, and with the
4D BSE,
\begin{equation} i(2\pi)^{4}\Delta_{1}\Delta_{2}\Phi(P,q)=\int
d^{4}qK(q,q^{\prime})\Phi(P,q^{\prime}),\label{eq:2.1}
\end{equation}
where $\Delta_{1,2}=m_{1,2}^2+p_{1,2}^2$are the inverse propagators,
and $m_{1,2}$ are (effective) constituent masses of quarks. The
4-momenta of the quark and anti-quark, $p_{1,2}$, are related to the
internal 4-momentum $q_{\mu}$ and total momentum $P_{\mu}$ of hadron of
mass $M$ as $p_{1,2}{}_{\mu}=\widehat{m}_{1,2}P_{\mu}\pm q_{\mu},$
where $\widehat{m}_{1,2} =[1\pm(m_1^2-m_2^2)/M^2]/2$ are the
Wightman-Garding (WG) definitions of masses of individual quarks. Now it
is convenient to express the internal momentum of the hadron $q$ as the
sum of two parts, the transverse component,
$\hat{q}_{\mu}=q_{\mu}-(q\cdot P)/P^2$ which is orthogonal to total hadron
momentum $P$ (ie. $\widehat{q}\cdot P=0$ regardless of whether the individual
quarks are on-shell or off-shell), and the longitudinal component,
$\sigma P_{\mu} = P_{\mu} (q\cdot P)/P^2$, which is parallel to P. Thus
we can decompose $q_{\mu}$ as, $q_\mu=(\widehat{q},Md\sigma)$, where
the transverse component, $\widehat{q}$ is an effective 3D vector, while
the longitudinal component, $Md\sigma$ plays the role of the fourth
component and is like the time component. To obtain the 3D BSE and the
Hadron-quark vertex, use an Ansatz on the BS kernel $K$ in Eq.
(\ref{eq:2.1}) which is assumed to depend on the 3D variables
$\hat{q}_{\mu}$, $\hat{q}_{\mu}^{\prime}$ \cite{mitra92} i.e.
\begin{equation}
K(q,q^{\prime})=K(\hat{q},\hat{q}^{\prime}).
\end{equation}
Hence, the longitudinal component, $\sigma P_{\mu}$ of
$q_{\mu}$, does not appear in the form $K(\hat{q},\hat{q}^{\prime})$ of
the kernel. Defining $\phi(\hat{q})$ as the 3D wave function,
\begin{equation} \
\phi(\hat{q})=\int\limits_{-\infty}^{+\infty}{Md\sigma\Phi(P,q)},
\end{equation} Integrating Eq.(1), and making use of Eqs.(2-3), we
obtain the 3D Salpeter equation, 
\begin{equation} \
(2\pi)^{3}D(\widehat{q})\phi(\widehat{q})=\int
d^{3}\widehat{q}'K(\widehat{q},\widehat{q}')\phi(\widehat{q}'),
\end{equation} 
where $D(\hat{q})$ is the 3D denominator function defined as 
\cite{bhatnagar06,bhatnagar11,elias11},
\begin{eqnarray}
\nonumber\frac{1}{D(\widehat{q})}&=&\frac{1}{2\pi
i}\int\limits_{-\infty}^{+\infty}\frac{
Md\sigma}{\Delta_1\Delta_2}=\displaystyle\frac{\displaystyle\frac{1}{2\omega_1}
+ \frac{1}{2\omega_2}} {(\omega_1+\omega_2)^2-M^2},\\
\omega_{1,2}^2&=&m_{1,2}^2+\hat{q}^2,
\end{eqnarray}
whose value given above is obtained by evaluating the contour
integration over inverse quark propagators in the complex $\sigma$-plane
by noting their corresponding pole positions (for details
see\cite{bhatnagar06,bhatnagar11}). which is used for making contact
with mass spectra of $q\overline{q}$ mesons.

Further, making use of Eq.(2) and (3) on RHS of Eq.(1), we get,
\begin{equation}
\ i(2\pi)^{4}\Delta_{1}\Delta_{2}\Phi(P,q)=\int d^{3}\widehat{q}'K(\widehat{q},\widehat{q}')\phi(\widehat{q}').
\end{equation}
From equality of RHS of Eq.(4) and (6), we see that an exact
interconnection between 3D and 4D BS wave functions is thus brought out.
The 4D Hadron-quark vertex function for scalar quarks under CIA can be
identified as:
\bigskip
\begin{equation}
\Delta_{1}\Delta_{2}\Phi(P,q)=\frac{D(\hat{q})\phi(\hat{q})}{2\pi i}
\equiv\Gamma.\label{eq:2.7}
\end{equation}

Now for fermionic quarks, the 4D BSE under gluonic which is akin to
vector type interaction kernel with a 3D support can be written as:
\begin{eqnarray}
\nonumber\ i(2\pi)^{4}\Psi(P,q)&=&S_{F1}(p_{1})S_{F2}(-p_{2})\int d^{4}q'K(\widehat{q},\widehat{q}')\Psi(P,q');\\
K(\widehat{q},\widehat{q}')&=&F_{12}i\gamma_{\mu}^{(1)}\gamma_{\mu}^{(2)}V(\widehat{q},\widehat{q}')
\end{eqnarray}

Here, $F_{12}$ is the color factor, 
$(\bm{\lambda}_1/2)\cdot(\bm{\lambda}_2/2)$ and the potential $V$ involves the 
scalar structure of the gluon propagator in the perturbative (o.g.e.) as 
well as the non-perturbative (confinement) regimes. The full structure 
of $V$ is \cite{mitra01}:

\begin{eqnarray}
\nonumber K(q,q^{\prime})&=&K(\hat{q},\hat{q}^{\prime})\\
\nonumber K(\widehat{q},\widehat{q'})&=&\left(\frac{1}{2}\bm{\lambda}_1\cdot\frac{1}{2}\bm{\lambda}_2\right)\gamma^{(1)}_{\mu}\gamma^{(2)}_{\mu}V(\widehat{q}-\widehat{q'})\\
\nonumber V(\hat{q}-\hat{q'})&=&V_{OGE}+V_C\\
\nonumber V_{OGE}&=&\frac{4\pi\alpha_{s}(Q^{2})}{(\widehat{q}-\widehat{q}')^{2}}\\
\nonumber V_C&=&\frac{3\omega_{q\bar{q}}^2}{4}\int d^3\bm{r}f(r)
e^{i(\hat{q}-\hat{q'})\cdot\bm{r}}\\
f(r)&=&
\frac{r^2}{(1+4a_0\hat{m_1}\hat{m_2}M^2r^2)^{1/2}}
-\frac{C_0}{\omega_0^2},
\end{eqnarray}
which is taken as one-gluon-exchange like as regards color
[$(\bm{\lambda}^{(1)}/2)\cdot(\bm{\lambda}^{(2)}/2)$] and spin ($
\gamma_\mu^{(1)}\gamma_\mu^{(2)}$) dependence. The scalar function
$V(q-q')$ is a sum of one-gluon exchange $V_{OGE}$ and a confining term
$V_C$. This confining term simulates an effect of an almost linear
confinement ($\sim r$) for heavy quark ($c,b$) sector, while retaining
harmonic form ($\sim r^2$) for light quark ($u,d$) sector as is believed to be true for QCD.
\begin{eqnarray}
\nonumber\ \omega_{q\overline{q}}^2&=&4\widehat{m}_1\widehat{m}_2M_>
\omega_0^2\alpha_S(M_>^2),\\
M_>&=&\mbox{Max}(M,m_1+m_2).
\end{eqnarray}
The values of basic constants are: $C_0=0.29, \omega_0=0.158$ GeV,
$\Lambda=0.200$ GeV, $m_{u,d}=0.265$ GeV, $m_s=0.415$ GeV, $m_c=1.530$
GeV and $m_b=4.900$ GeV \cite{mitra01,mitra99,bhatnagar11}. However the form of
BSE in Eq.(8) is not convenient to use in practice since
Dirac matrices lead to several coupled integral equations. However a
considerable simplification is effected by expressing them in
Gordon-reduced form which is permissible on the mass shells of quarks
(ie. on the surface $P\cdot q=0$). The Gordon reduced BSE form of the
fermionic BSE can be written as \cite{mitra01}:
\begin{eqnarray}
\nonumber\ \Delta_{1}\Delta_{2}\Phi(P,q)&=&-i(2\pi)^{4}\int d^{4}q'\widetilde{K}_{12}(\widehat{q},\widehat{q}')\Phi(P,q'),\\
\widetilde{K}_{12}(\widehat{q},\widehat{q}')&=&F_{12}V_{\mu}^{(1)}V_{\mu}^{(2)}V(\widehat{q},\widehat{q}')
\end{eqnarray}
where the connection between $\Psi$ and $\Phi$ (whose structure is
identical as $\Phi$ in case of scalar quarks in Eq.(1)) is,
\begin{eqnarray}
\nonumber\Psi(P,q)&=&(m_1-i\gamma^{(1)}\cdot p_1)(m_{2}+i\gamma^{(2)}\cdot p_2)\Phi(P,q),\\
\nonumber V_{\mu}^{(1,2)}&=&\pm 2m_{1,2}\gamma_{\mu}^{1,2},\\
V_{\mu}^{(1,2)}&=&(p_{1,2}+p'_{1,2})_{\mu}+i\sigma_{\mu\nu}^{(1,2)}(p_{1,2}+p'_{1,2})_{\nu}.
\end{eqnarray}

Now to reduce the above BSE to the 3D form, all time-like components 
$\sigma,\sigma'$ of momenta in $V_{\mu}^{(1)}V_{\mu}^{(2)}$ on RHS of 
Eq.(9) are replaced by their on-shell values giving us the 3D form, 
$\bm{V}_1\cdot\bm{V}_2$. Thus
\begin{eqnarray}
&&\nonumber\ V_{\mu}^{(1)}V_{\mu}^{(2)}\Rightarrow\bm{V}_1\cdot\bm{V}_2=-4\widehat{m}_{1}\widehat{m}_{2}M^{2}-(\widehat{q}-\widehat{q}')^{2}-2(\widehat{m}_{1}-\widehat{m}_{2})P\cdot(\widehat{q}+\widehat{q}')\\&&
\nonumber\ -i(2\widehat{m}_1P+\widehat{q}+\widehat{q}')_i\sigma_{ij}^{(2)}(\widehat{q}-\widehat{q}')+i(2\widehat{m}_2P-\widehat{q}-\widehat{q}')_i\sigma_{ij}^{(1)}(\widehat{q}-\widehat{q}')_j
 +\sigma_{ij}^{(1)}\sigma_{ij}^{(2)}.
\end{eqnarray}

The 3D form of BSE then works out as \cite{mitra99}:
\begin{eqnarray}
\nonumber D(\widehat{q})\phi(\widehat{q})&=&\omega_{q\overline{q}}^2\widetilde{D}(\widehat{q})\phi(\widehat{q}),\\
\nonumber \widetilde{D}(\widehat{q})&=&4\widehat{m}_1\widehat{m}_2M^2(\bm{\nabla}^2+C_0/\omega_0^2)+4\widehat{q}^2\bm{\nabla}^2+8\widehat{q}\cdot\bm{\nabla}\\
&& +18-8\bm{J}\cdot\bm{S}+\frac{4C_0}{\omega_0^{2}}\widehat{q}^2.
\end{eqnarray}
This is reducible to the equation for a 3D harmonic oscillator with
coefficients depending on the mass $M$ and total quantum number $N$. The
ground state wave functions \cite{mitra01,mitra99,bhatnagar06} deducible
from this equation have gaussian structure and are expressed as:
\begin{equation}
\ \phi(\widehat{q})=e^{-\widehat{q}^2/2\beta^2}.
\end{equation}
and is appropriate for making contact with O(3)-like mass spectrum (for
details see \cite{mitra01}). It is to be noted that this 3D BSE (in
Eq.(14)) which is responsible for determination of mass spectra of
mesons in CIA is formally equivalent (see \cite{mitra01,mitra99,bhatnagar05})
to the corresponding mass spectral equation deduced earlier using
Null-Plane Approximation (NPA)\cite{mitra90}. Thus the mass spectral predications for $q\bar{q}$ systems in BSE under CIA are identical to the corresponding mass spectral predictions for these systems in
BSE under NPA \cite{mitra90} (see \cite{mitra01,mitra99} for details).

We further wish to mention that a similar form for ground state wave
function in harmonic oscillator basis using variational arguments has
been used in \cite{arndt99}. In ground state wave function
$\phi(\widehat{q})$ in Eq.(15), $\beta$ is the inverse range parameter
which incorporates the content of BS dynamics and is dependent on the
input kernel $K(q,q^\prime)$ (for details see
\cite{mitra01,bhatnagar06,bhatnagar11}) The structure of inverse range
parameter $\beta$ in wave functions $\phi(\widehat{q})$ is given as
\cite{mitra01,bhatnagar06,bhatnagar11,elias11}:
\begin{eqnarray}
\nonumber\beta^2&=&\left(\frac{2}{\gamma^2}\widehat{m}_1\widehat{m}_2M\omega_{q\bar{q}}^2\right)^{1/2},\\
\nonumber\gamma^2&=&1-\frac{2\omega_{q\bar{q}}^2C_0}{M_>\omega_0^2},\\
\nonumber\omega_{q\overline{q}}^2&=&4\widehat{m}_1\widehat{m}_2M_>
\omega_0^2\alpha_S(M_>^2),\\
\ M_>&=&\mbox{Max}(M,m_1+m_2).
\end{eqnarray}

\subsection{Dirac structure of Hadron-quark vertex function for P-mesons
in BSE with power counting scheme}

Thus, for fermionic quarks, the full 4D BS wave function can be written
as

\begin{equation}
\Psi(P,q)=S_{F}(p_1)\Gamma S_{F}(-p_2),
\end{equation}
where the 4D hadron-quark vertex function is
\cite{mitra01,bhatnagar06,bhatnagar06,bhatnagar11} i.e.
\begin{equation}
\Gamma=\frac{1}{2\pi i}(\Omega\cdot\varepsilon) N_{V}D(\widehat{q})\phi(\widehat{q}).
\end{equation}
The 4D hadron-quark vertex $\Gamma$ in the above equation, satisfies a 4D BSE
with a natural off-shell extension over the entire 4D space (due to the
positive definiteness of the quantity $\hat{q}^{2}=q^{2}-(q\cdot
P)^{2}/P^{2}$ throughout the entire 4D space) and thus provides a fully
Lorentz-invariant basis for evaluation of various transition amplitudes
through various quark loop diagrams. $N_{V}$ in the above equation is the 4D BS normalizer.

In the hadron-quark vertex function, $\Gamma$ above,
($\Omega\cdot\varepsilon$) contains the relevant Dirac structures
\cite{smith69} which makes $\Gamma$ a $4\times4$ matrix in
the spinor space. In our model, the relevant Dirac structures in
$\Gamma$ are incorporated in accordance with our recently
proposed power counting rule \cite{bhatnagar06,shi07} which identifies
the leading order (LO) Dirac structures from the next-to-leading order
(NLO) Dirac structures and for a vector meson is expressed as:
\begin{eqnarray}
\nonumber
(\Omega\cdot\varepsilon)&=&i(\gamma\cdot\varepsilon)A_{0}+(\gamma\cdot\varepsilon)(\gamma\cdot P)
\frac{A_1}{M}
+[q\cdot\varepsilon-(\gamma\cdot\varepsilon)(\gamma\cdot q)]\frac{A_2}{M}\\&&
\nonumber+i\frac{A_3}{M^2}[(\gamma\cdot\varepsilon)(\gamma\cdot P)(\gamma\cdot q) -
(\gamma\cdot\varepsilon)(\gamma\cdot q)(\gamma\cdot P)+2i(q\cdot\varepsilon)(\gamma\cdot P)]\\&&
+(q\cdot\varepsilon)\frac{A_4}{M}+i(q\cdot\varepsilon)(\gamma\cdot P)\frac{A_5}{M^2}
- i(q\cdot\varepsilon)(\gamma\cdot q)\frac{A_6}{M^2}
+(q\cdot\varepsilon)[(\gamma\cdot P)(\gamma\cdot q)-(\gamma\cdot q)(\gamma\cdot P)]\frac{A_7}{M^3},
\end{eqnarray}
where $A_i(i=0,...,7)$ are taken as the eight dimensionless constant
coefficients to be determined. But since we take constituent quark
masses, where the quark mass, $M$ is approximately half the hadron mass,
$M$, we can use the ansatz,
\begin{equation}
\ q\ll P\sim M
\end{equation}
in the rest frame of the hadron. Then each of the eight terms in the
above equation receives suppression by different powers of $1/M$. Thus
we can arrange these terms as an expansion in powers of $O(1/M)$. We can
then see that in the expansion of $\Gamma\cdot\varepsilon$ that the
structures associated with the coefficients, $A_{0},A_1$ have
magnitudes, $O(1/M^0)$ and are of leading order (LO). Those with $A_2,
A_3, A_4, A_5$ are $O(1/M^1)$ and are next-to-leading order (NLO), while
those with $A_6,A_7$ are $O(1/M^2)$ and are NNLO. This naive power
counting rule suggests that the maximum contribution to calculation of
any vector meson observable should come from Dirac structures
$\gamma\cdot\varepsilon$ and $(\gamma\cdot\varepsilon)(\gamma\cdot P)/M$
associated with coefficients, $A_0$ and $A_1$ respectively, followed by
the higher order Dirac structures associated with the other four
coefficients, $A_2,A_3,A_4,A_5$ and then by Dirac structures associated
with coefficients, $A_6,A_7$.

In this work we try to calculate the decay constants using LO and NLO
Dirac structures since we wish to calculate the decay constants up to
next-to-leading orders. Thus we take the form of the hadron-quark vertex
function as,
\begin{eqnarray}
\nonumber\ \Gamma&=&\left\{i(\gamma\cdot\varepsilon)A_{0} +
(\gamma\cdot\varepsilon)(\gamma\cdot P)\frac{A_1}{M}\right.
+\left[q\cdot\varepsilon-(\gamma\cdot\varepsilon)(\gamma\cdot q)\right]\frac{A_2}{M}\\&&
\nonumber+i\left[(\gamma\cdot\varepsilon)(\gamma\cdot P)(\gamma\cdot q) -
(\gamma\cdot\varepsilon)(\gamma\cdot q)(\gamma\cdot P)+2i(q\cdot\varepsilon)(\gamma\cdot P)\right]
\frac{A_3}{M^2}\\&&
\left.+(q\cdot\varepsilon)\frac{A_4}{M}+i(q\cdot\varepsilon)(\gamma\cdot P)
\frac{A_5}{M^2}\right\}\frac{1}{2\pi i}N_VD(\widehat{q})\phi(\widehat{q}).
\end{eqnarray}
In the above equation, $N_{V}$ is the 4D BS normalizer for ground state
vector meson with internal momenta $q$, and is worked out in the
framework of Covariant Instantaneous Ansatz (CIA) to give explicit
covariance to the full fledged 4D BS wave functions, $\Psi(P,q)$ and
hence to the Hadron-quark vertex function, $\Gamma$, employed
for calculation of decay constants. In the structure of
$\Gamma$ above, $\phi(\widehat{q})$ is the ground state 3D BS
wave function for vector meson with internal momenta $q$ and is given in Eq.(15).

\section{Electromagnetic decays of vector mesons through the process
$\bf V\rightarrow\gamma*\rightarrow e^++e^-$}
\label{sec:methodology}
\subsection{Transition amplitude}
The vector meson decay proceeds through the quark-loop diagram shown
below (See Fig.1).

\begin{figure}[bt]
\begin{center}
\includegraphics[width=0.8\linewidth,height=0.3\linewidth]{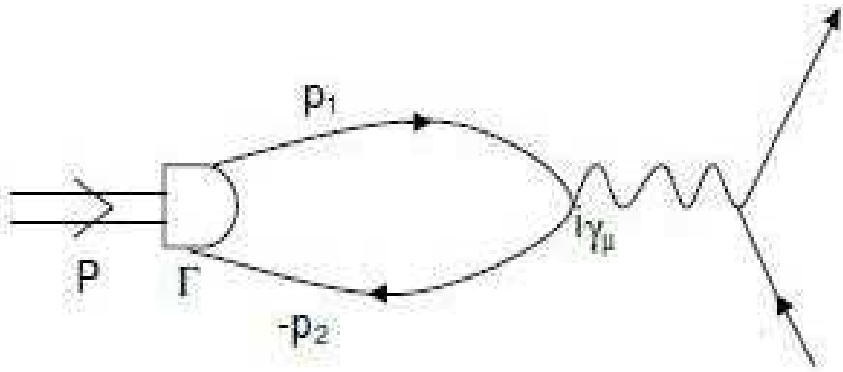}
\caption{Quark loop diagram for $V\rightarrow\gamma^*\rightarrow
e^++e^-$ showing the coupling of electromagnetic current to the quark
loop.}
\label{fig:quarkloop}
\end{center}
\end{figure}

The coupling of a vector meson of momentum $P$ and polarization
$\varepsilon_{\mu}$ to the photon is expressed via dimensionless
coupling constant $g_V$ which can be described by the matrix element,
\begin{equation}
\ \frac{M^2}{g_V}\varepsilon_{\mu}(P)=\langle0\vert\overline{Q}\widehat{\Theta}\gamma_{\mu}Q\vert V(P)\rangle
\end{equation}
(where $Q$ is the flavour multiplet of quark field and
$\widehat{\Theta}$ is the quark electromagnetic charge operator) which
can in turn be expressed as a loop integral,
\begin{equation}
\ \frac{M^2}{g_V}\varepsilon_\mu(P)=\sqrt3e_Q\int d^4q Tr[\Psi_V(P,q)i\gamma_\mu].
\end{equation}
Here $e_Q$ arises from the flavour configuration of individual vector
mesons and has values: $e_Q=1/\sqrt{2},1/3,1/\sqrt{18},2/3$ and $1/3$
for $\rho,\phi,\omega,\psi$ and $Y$ respectively, and the polarization
vector $\varepsilon_\mu$ of V-meson satisfies, $\varepsilon\cdot P=0$.
Defining leptonic decay constant,$f_{V}$ as,
$f_{V}=M/(e_{Q}g_{V})$ \cite{bhatnagar06}, we can express
\begin{equation}
\ f_{V}\varepsilon_{\mu}(P)=\frac{\sqrt{3}}{M}\int d^{4}q Tr\left[\Psi_{V}(P,q)i\gamma_{\mu}\right]
\end{equation}
Plugging $\Psi_{V}(P,q)$, which involves the structure of hadron-quark
vertex function in Eq.~(2.8) flanked by the Dirac propagators of the two
quarks as in Eq.~(2.3), into the above equation, evaluating trace over the
gamma matrices and noting that only the components of terms on the right
hand side in the direction of $\varepsilon_\mu$ will contribute to the
integral, we multiply both sides of the above integral by $P_\mu/M^2$,
we can then express the leptonic decay constant $f_V$ as,
\begin{equation}
\ f_V=f_V^0+f_V^1+f_V^2+f_V^3+f_V^4+f_V^5
\end{equation}
where $f_V^0, f_V^1,...,f_V^5$ are contributions to $f_V$ from the six
Dirac structures associated with $A_0,A_1,...,A_5$ in the expression
for hadron-quark vertex function $\Gamma(\widehat{q})$ and are expressed
analytically in terms of $d\sigma$ integrations over the poles
$\Delta_{1,2}$ of the quark propagators as:
\begin{eqnarray}
\nonumber\ f_V^0&=&\sqrt{3}N_V\frac{A_0}{M}\int d^3\widehat{q}D(\widehat{q})\phi(\widehat{q})\int \frac{Md\sigma}{2\pi i\Delta_1\Delta_2}4\left[\left(\frac{M^2}{6}+\frac{2}{3}m^2\right)+\frac{\Delta_1+\Delta_2}{6}\right]\\
\nonumber\ f_V^1&=&\sqrt{3}N_V\frac{A_1}{M}\int d^3\widehat{q}D(\widehat{q})\phi(\widehat{q})\int\frac{Md\sigma}{2\pi i\Delta_1\Delta_2}(-4mM)\\
\nonumber\ f_V^2&=&\sqrt{3}N_V\frac{A_2}{M}\int d^3\widehat{q}D(\widehat{q})\phi(\widehat{q})\int\frac{Md\sigma}{2\pi i\Delta_1\Delta_2}\left[-\frac{4}{3}m(\Delta_1-\Delta_2)\right]\\
\nonumber\ f_V^3&=&\sqrt{3}N_V\frac{A_3}{M}\int d^3\widehat{q}D(\widehat{q})\phi(\widehat{q})\int\frac{Md\sigma}{2\pi i \Delta_1\Delta_2}\left[-\frac{8}{3}(\Delta_1+\Delta_2)+\left(\frac{16}{3}m^2-\frac{4}{3}M^2\right)\right]\\
\nonumber\ f_V^4&=&\sqrt{3}N_V\frac{A_4}{M}\int d^3\widehat{q}D(\widehat{q)}\phi(\widehat{q})\int\frac{Md\sigma}{2\pi i \Delta_1\Delta_2}\left[-\frac{2m}{3M}(\Delta_1+\Delta_2)+\left(\frac{4m^3}{3M}-\frac{1}{3}mM\right)\right]\\
\ f_V^5&=&\sqrt{3}N_V\frac{A_5}{M}\int d^3\widehat{q}D(\widehat{q)}\phi(\widehat{q})\int\frac{Md\sigma}{2\pi i \Delta_1\Delta_2}\left[\left(\frac{4}{3}m^2-\frac{2}{3}M^2\right)(\Delta_1-\Delta_2)\right].
\end{eqnarray}
In deriving the above expressions, we have made use of the following
relation showing the normalization over the polarization vector,
$\varepsilon(P)$ for $V$- meson of 4-momentum, $P$ as,
\begin{equation}
\ \varepsilon_\mu\varepsilon_\nu=\frac{1}{3}\left(\delta_{\mu\nu}+\frac{P_\mu P_\nu}{M^2}\right),
\end{equation}
We made use of the above relation to express the quantities involving
dot products of $\varepsilon$ with various momenta like,
$(p_1\cdot\varepsilon)(p_2\cdot\varepsilon)$,
$(p_1\cdot\varepsilon)(q\cdot\varepsilon)$ and
$(p_2\cdot\varepsilon)(q\cdot\varepsilon)$ in terms of dot products of momenta
as,
\begin{eqnarray}
&&\nonumber\ (p_1\cdot\varepsilon)(p_2\cdot\varepsilon)=\frac{1}{3}p_1\cdot p_2-\frac{(p_1\cdot P)(p_2\cdot P)}{3M^2}\\&&
\nonumber\ (p_1\cdot\varepsilon)(q\cdot\varepsilon)= \frac{1}{6}(p_1^2-p_1\cdot p_2)-\frac{(p_1\cdot P)(p_1\cdot P-p_2\cdot P)}{6M^2}\\&&
\ (p_2\cdot\varepsilon)(q\cdot\varepsilon)=\frac{1}{6}(p_1\cdot p_2-p_2^2)-\frac{(p_2\cdot P)(p_1\cdot P-p_2\cdot P)}{6M^2}
\end{eqnarray}

These dot products of momenta were in turn expressible
in terms of the inverse propagators, $\Delta_{1,2}$ as:
\begin{eqnarray}
\nonumber p_1\cdot P&=&\frac{1}{2}(\Delta_1-\Delta_2-M^2)\\
\nonumber p_2\cdot P&=&\frac{1}{2}(-\Delta_1+\Delta_2-M^2)\\
\nonumber p_1\cdot p_2&=&m^2-\frac{1}{2}(\Delta_1+\Delta_2+M^2)\\
\nonumber p_{1,2}^2&=&\Delta_{1,2}-m^2\\
\nonumber p_1\cdot q&=&\frac{3}{4}\Delta_1+\frac{1}{4}(\Delta_2+M^2)-m^2\\
\nonumber p_2\cdot q&=&-\frac{3}{4}\Delta_2-\frac{1}{4}(\Delta_1+M^2)+m^2\\
P\cdot q&=&\frac{1}{2}(\Delta_1-\Delta_2)
\end{eqnarray}

Thus all expressions for $f_{V}^{i}$ above were expressible in terms of
$\Delta_{1,2}$. Then carrying out integrations over the off-shell
variable, $d\sigma$ by the method of contour integrations by noting the
pole positions in the complex $\sigma-$ plane:
\begin{eqnarray}
\nonumber \Delta_1&=&0\Rightarrow\sigma_1^\pm=\pm \frac{\omega_1}{M}-\widehat{m}_1\mp i\epsilon,\\
\nonumber \Delta_2&=&0\Rightarrow\sigma_2^\mp=\mp\frac{\omega_2}{M}-\widehat{m}_2\pm i\epsilon,\\
\nonumber \omega_1^2&=&\omega_2^2=m^2+\widehat{q}^2\\
\widehat{m}_1&=&\widehat{m}_2=\frac{1}{2}
\end{eqnarray}

we get, the following integrals:
\begin{eqnarray}
\nonumber\int\frac{Md\sigma}{\Delta_1\Delta_2}(\Delta_1+\Delta_2)&=&D_0(\widehat{q});\\
\int\frac{Md\sigma}{\Delta_1\Delta_2}&=&\frac{1}{D(\widehat{q})},
\end{eqnarray}

where,
\begin{eqnarray}
\nonumber D(\widehat{q})&=&\omega D_0(\widehat{q}),\\
D_0(\widehat{q})&=&4\omega^2-M^2.
\end{eqnarray}
Thus we can express the various components $f_V^i$ $(i=0,...,5)$ of
$f_{V}$ in Eq.~(3.5) as,

\begin{eqnarray}
&&\nonumber\ f_V^0=\sqrt{3}N_V\frac{A_0}{M}\int d^3\widehat{q}\phi(\widehat{q})4\left[\frac{M^2}{6}+\frac{2}{3}m^2+\frac{D_0(\widehat{q})}6\right]\\&&
\nonumber\ f_V^1=\sqrt{3}N_V\frac{A_1}{M}\int d^3\widehat{q}\phi(\widehat{q})(-4mM)\\&&
\nonumber\ f_V^2=0\\&&
\nonumber\ f_V^3=\sqrt{3}N_V\frac{A_3}{M}\int d^3\widehat{q}\phi(\widehat{q})\left[-\frac{8}{3}D_{0}(\widehat{q})+\frac{16}{3}m^2-\frac{4}{3}M^2\right]\\&&
\nonumber\ f_V^4=\sqrt{3}N_V\frac{A_4}{M}\int d^3\widehat{q}\phi(\widehat{q})\left[-\frac{2m}{3M}D_0(\widehat{q})+\frac{4m^3}{3M}-\frac{1}{3}mM\right]\\&&
\ f_V^5=0
\end{eqnarray}

Note that the components, $f_V^2$ and $f_V^5$ are $0$ on account of
equal mass kinematics.

Note that each of the $f_V^i$ involves the BS normalizer $N_V$. This is
evaluated using the current conservation
condition:\cite{bhatnagar06,bhatnagar11,elias11},
\begin{equation}
2iP_\mu=(2\pi)^4\int d^4q \mbox{Tr}\left\{\overline{\Psi}(P,q)\left[\frac{\partial}{\partial P_{\mu}}S_{F}^{-1}(p_1)\right]\Psi(P,q)S_{F}^{-1}(-p_2)\right\} + (1\rightleftharpoons2)
\end{equation}
Putting BS wave function $\Psi(P,q)$ for a vector meson in Eqs.~(2.3)
and (2.8) in the above equation, carrying out derivatives of inverse
quark propagators of constituent quarks with respect to total hadron
momentum $P_\mu$, evaluating trace over products of gamma matrices,
following usual steps, and multiplying both sides of the equation by
$P_\mu/(-M^2)$ to extract out the normalizer $N_V$ from the above
equation, we then express the above equation in terms of the integration
variables $\widehat{q}$ and $\sigma$. Noting that the 4D volume element
$d^4q =d^3\widehat{q}Md\sigma$, we then perform the contour integration
in the complex $\sigma$- plane by making use of the corresponding pole
positions. For details of these mathematical steps involved in the
calculations of BS normalizers for vector and pseudoscalar mesons see
\cite{bhatnagar06,bhatnagar11}, where in the present calculation, we
take both, the leading order (LO) as well as the next-to-leading order
(NLO) Dirac structures for vector mesons in their respective 4D BS wave
functions $\Psi(P,q)$. Then integration over the variable $\widehat{q}$
is finally performed to extract out the numerical results for $N_V$ for
different vector mesons. The calculation of $N_V$ is quite complex due
to the $6$ Dirac structures involved in the calculation. The structure
of the normalizer is of the form,
\begin{equation}
N_V^{-2}=i(2\pi)^2\int d^3\widehat{q}D^2(\widehat{q})\phi^2(\widehat{q})
\displaystyle\sum\limits_{ij}A_i A_jI_{ij}(m,M,\widehat{q},S).
\label{eq:Nvm2}
\end{equation}
Here, $A=(A_0,A_1,A_2,A_3,A_4,A_5)$ are functions of parameters $A_i$,
$S=(R,D_1,D_2,D_{11},$ $D_{12},D_{22})$, where the $S_i$ are the results
of integrations over off-shell parameter $\sigma$ whose results are
given in Eq.~(\ref{eq:theSi}). The $I_{ij}(m,M,\widehat{q},R)$ are
involved functions of $m,M,\widehat{q}$ and $R$. Explicit expressions
are listed below in Eq.~(\ref{eq:I20}):
\begin{eqnarray*}
I_{00}&=&\frac{1}{3}\left[\frac{1}{M^2}\left(1-12\frac{m^2}{M^2}\right)D_{11}+\frac{3}{M^2}\left(-1+\frac{m^2}{M^2}\right)D_{12}+\frac{1}{M^2}(7m^4-3m^2M^2)R\right.\\
&\phantom{=}&\left.+\frac{1}{M^2}\left(M^2+\frac{13m^4}{M^2}-5m^2\right)D_1-\frac{13m^4D_2}{M^4}+\frac{9m^2D_{22}}{M^4}\right]\\
I_{11}&=&\frac{2}{3}\left[-\frac{4D_{11}}{M^2}+\frac{3D_{12}}{M^2}-\frac{4D_{22}}{M^2}-\frac{8M^2D_1}{M^2}+\frac{(8m^2+M^2)D_2}{M^2}+2(4m^2+M^2)R\right]\\
I_{22}&=&\frac{2}{3}\left[\frac{D_{11}}{M^2}+\frac{3D_{12}}{M^2}-\frac{4D_{22}}{M^2}-\frac{8m^2D_1}{M^2}+\frac{(8m^2+M^2)D_2}{M^2}+2(4m^2+M^2)R\right]\\
I_{33}&=&\frac{1}{3}\left[\frac{D_{11}}{M^2}\left(\frac{-6m^4}{M^4}+\frac{9m^2}{M^2}-\frac{22m^3}{M^3}-\frac{55m^2}{2M^2}+13\right)\right.\\
&\phantom{=}& +\frac{D_{12}}{M^2}\left(\frac{18m^4}{M^4}-\frac{7m^2}{M^2}+\frac{70m^2}{M^2}+\frac{12m^3}{M^3}-4\right)\\
&\phantom{=}& +\frac{D_{22}}{M^2}\left(\frac{9m^4}{M^4}-\frac{67m^2}{M^2}+\frac{10m^3}{M^3}+35-\frac{9m^2}{M^2}\right)\\
&\phantom{=}& +\frac{D_1}{M^2}\left(\frac{29}{2}M^2-13m^2+\frac{12m^4}{M^2}-\frac{3m^3}{M}-\frac{48m^4}{M^2}+\frac{16m^6}{M^4}-\frac{36m^5}{M^3}\right)\\
&\phantom{=}& +\frac{D_2}{M^2}\left(22M^2-7m^2-\frac{12m^4}{M^2}+\frac{3m^3}{M}+\frac{48m^4}{M^2}-\frac{16m^6}{M^4}+\frac{4m^5}{M^3}\right)\\
&\phantom{=}&\left.+\frac{R}{M^2}\left(-m^2M^2-14m^3M+76m^4+16mM^3+6M^4+\frac{5}{2}m^4\right)\right]\\
I_{44}&=&\frac{1}{3}\left[\frac{D_{11}}{M^2}\left(\frac{4m^2}{M^2}+1\right)+\frac{7D_{12}}{2M^2}+\frac{D_{22}}{M^2}\left(\frac{1}{6}-\frac{4m^2}{M^2}\right)\right.\\
&\phantom{=}& +\left(2-\frac{6m^4}{M^4}-\frac{m^5}{M^5}-\frac{6m^2}{M^2}-\frac{m^3}{M^3}\right)D_1+\left(\frac{1}{6}+\frac{m^5}{M^5}+\frac{6m^4}{M^4}-\frac{8m^2}{M^2}\right)D_2\\
&\phantom{=}&\left. +\left(-4m^2+\frac{8m^4}{M^2}+\frac{m^5}{M^3}+\frac{M^2}{3}\right)R\right]\\
I_{55}&=&\frac{1}{3}\left[-\frac{3}{2M^2}\left(\frac{m^4}{3M^4}+\frac{11m^2}{3M^2}+\frac{3}{2}\right)D_{12}+\frac{D_1}{M^2}\left(-M^2+\frac{3m^4}{2M^2}+\frac{7m^5}{2M^3}+\frac{17m^2}{4}\right)\right.\\
&\phantom{=}&\left.+\frac{D_2}{M^2}\left(-\frac{M^2}{4}-\frac{3m^5}{2M^3}-\frac{7m^4}{2M^2}+\frac{15m^2}{4}\right)+\frac{R}{M^2}\left(-\frac{3M^4}{2}+\frac{9m^2M^2}{4}+\frac{7m^4}{2}\right)R\right]\\
I_{01}&=&\left[-\frac{4mD_{11}}{M^3}-\frac{8mD_{22}}{M^3}+\frac{4mD_{12}}{M^3}-\frac{4D_1}{3M}\left(\frac{4m^3}{M^2}+m\right)\right.\\&&
\left.+\frac{4D_2}{M}\left(\frac{4m^3}{3M^2}-m\right)+\left(\frac{16m^3}{3M}-\frac{4mM}{3}R\right)\right]\\
I_{02}&=&\left[\frac{8m}{M}(D_1-D_2)-8mMR\right];\\
I_{03}&=&\left[-\frac{2m}{M^3}D_{11}+\frac{4m}{3M^3}D_{12}+\frac{2m}{M^3}D_{22}+\frac{8D_1}{3M}(\frac{m^3}{M^2}-m)+\frac{3m^3}{3M^3}D_2\right.\\&&
\left.+(\frac{8m^3}{3M}-\frac{2mM}{3})R\right]\\
I_{04}&=&\left[-\frac{7m^2D_{11}}{3M^4}+\frac{D_{22}}{3M^2}\left(-\frac{2m^4}{M^4}-\frac{7m^2}{M^2}+8\right)+\frac{D_{12}}{3M^2}\left(\frac{2m^2}{M^2}-8\right)\right.\\&&
 \left.+\frac{D_1}{M^2}\left(-M^2-\frac{2m^4}{3M^2}+\frac{8m^2}{3}\right)+\frac{D_2}{M^2}\left(\frac{4M^2}{3}+\frac{2m^4}{3M^2}-\frac{8m^2}{3})+\frac{R}{M^2}(-m^2M^2-2m^4\right)\right]\\
I_{05}&=&\left[\frac{D_{11}}{M^2}\left(\frac{6m^2}{M^2}-\frac{34}{3}\right)+\frac{D_{22}}{3M^2}\left(\frac{26m^2}{M^2}-14\right)-\frac{D_{12}}{3M^2}\left(\frac{44m^2}{M^2}+20\right)\right.\\&&
 +\frac{D_1}{M^2}\left(-6M^2+\frac{8m^4}{3M^2}-12m^2\right)D_1+\frac{D_2}{M^2}\left(-\frac{22M^2}{3}-\frac{8m^4}{M^2}+\frac{44m^2}{3}\right)\\&&
 +\frac{R}{M^2}\left(14m^2M^2-\frac{8m^4}{3}-\frac{10M^4}{3}\right)\}\\
I_{12}&=&\left[\frac{D_{11}}{M^2}\left(\frac{m^2}{M^2}+\frac{2}{3}\right)+\frac{D_{22}}{M^2}\left(\frac{m^2}{M^2}-4\right)+\frac{D_{12}}{M^2}\left(-\frac{2m^2}{M^2}+\frac{10}{3}\right)\right.\\&&
 \left.+\frac{D_1}{M^2}\left(\frac{8M^2}{3}-\frac{2m^2}{3}\right)+\frac{D_2}{M^2}\left(-2M^2+\frac{2m^2}{3}\right)+\frac{5m^2R}{3}\right]\\
I_{13}&=&\left[\frac{D_{11}}{3M^3}\left(-\frac{4m^3}{M^2}-34m\right)+\frac{D_{22}}{3M^3}\left(-\frac{4m^3}{M^2}-26m\right)+\left(\frac{8m^2}{M^4}+\frac{20}{M^2}\right)D_{12}\right.\\&&
 +\frac{D_{22}}{M^3}\left(-\frac{16m^5}{3M^4}+\frac{8m^3}{3M^2}+28m\right)+\frac{D_2}{M}\left(\frac{16m^5}{3M^4}-\frac{8m^3}{3M^2}-\frac{46m}{3}\right)\\&&
 \left.+\left(-\frac{28m^4}{3M^2}-\frac{8m^3}{3M}-2mM\right)R\right]\\
I_{14}&=&\left[\frac{D_{11}}{M^2}\left(-\frac{m^4}{M^4}+\frac{4}{3}\right)+\frac{D_{22}}{M^2}\left(-\frac{m^2}{M^2}+2\right)+\frac{D_{12}}{M^2}\left(\frac{2m^2}{M^2}-\frac{10}{3}\right)\right.\\&&
 \left.+\frac{D_1}{M^2}\frac{13m^2}{3}-\frac{D_2}{M^2}\frac{5m^2}{3}+\left(-\frac{8m^2}{3}+\frac{2M^2}{3}\right)R\right]\\
I_{15}&=&\left[\frac{5m}{3M^3}D_{11}-\frac{m}{3M^3}D_{22}-\frac{4m}{3M^3}D_{12}+\frac{D_2}{M}\left(-\frac{2m^3}{M^2}+\frac{2m}{3}\right)D_2+\left(-\frac{2m^3}{M}+mM\right)R\right]\\
I_{23}&=&\left[\left(\frac{7m^3}{3M^5}+\frac{3m^2}{4M^4}+\frac{5m}{12M^3}\right)D_{11}+\left(\frac{7m^3}{3M^5}-\frac{3m^2}{6M^4}\right)D_{22}\right.\\&&
 +\left(-\frac{14m^3}{3M^5}+\frac{m}{3M^3}-\frac{3m^2}{4M^4}\right)D_{12}\\&&
 +\left(-\frac{7m^3}{6M^3}+\frac{m}{6M}-\frac{3m^4}{8M^4}+\frac{3m^2}{8M^2}\right)D_1+D_2\left(\frac{7m^3}{6M^3}-\frac{m}{M}+\frac{3m^4}{8M^4}\right)\\&&
 \left.+\left(\frac{5m^3}{6M}-\frac{mM}{4}+\frac{3m^4}{16M^2}\right)R\right];\\
I_{24}&=&\left[\frac{13m^3}{M}R+\frac{D_{11}}{M^3}\left(-\frac{3m^3}{2M^2}+\frac{7m}{3}\right)-\frac{D_{22}}{M^3}\left(\frac{m^3}{3M^2}+6m\right)+\frac{D_{12}}{M^3}\left(\frac{3m^3}{M^2}+\frac{11m}{3}\right)\right.\\&&
 \left.+\frac{D_1}{3M}\left(-\frac{25m^3}{M^2}-5m\right)+\frac{D_2}{3M}\left(\frac{25m^3}{M^2}-19m\right)\right]\\
I_{25}&=&\left[\frac{D_{11}}{M^2}\left(-\frac{3m^2}{2M^2}+\frac{4m^4}{M^4}+1\right)+\frac{D_{22}}{M^2}\left(-\frac{7m^2}{2M^2}-\frac{4m^4}{3M^4}+\frac{10}{3}\right)\right.\\&&
 +\frac{D_1}{M^2}\left(\frac{7M^2}{6}+\frac{m^4}{M^2}-10\frac{4m^2}{3}-\frac{8m^6}{3M^4}\right)+\frac{D_2}{M^2}\left(\frac{8M^2}{3}-\frac{m^4}{3M^2}+\frac{8m^6}{3M^4}-8m^2\right)\\&&
 \left.+\frac{D_1D_2}{M^2}\left(\frac{5m^2}{M^2}-\frac{8m^4}{3M^4}+2\right)+\left(-\frac{23m^2}{6}+\frac{19m^4}{3M^2}+\frac{8m^6}{3M^4}+\frac{2M^2}{3}\right)R\right]\\
I_{34}&=&\left[-\frac{m^3}{M^5}\left(D_{11}+D_{22}\right)+\frac{4m^3}{M^5}D_{12}\right]\\
I_{35}&=&\left[\frac{D_{11}}{M^2}\left(\frac{m^4}{M^4}+\frac{28m^2}{M^2}-\frac{16m^4}{M^4}-\frac{14}{3}\right)+\frac{D_{22}}{M^2}\left(\frac{m^2}{M^2}-\frac{8m^4}{3M^4}+\frac{16m^3}{M^3}-\frac{2}{3}\right)\right.\\&&
 +\frac{D_1}{M^2}\left(\frac{2M^2}{3}-\frac{10m^3}{M}-\frac{2m^4}{3M^2}+\frac{16m^6}{3M^4}+\frac{22m^2}{3}\right)+\frac{D_2}{M^2}\left(2M^2-\frac{12m^2}{3}+\frac{26m^4}{3M^2}-\frac{2m^4}{3M^2}-\frac{16m^6}{3M^4}\right)\\&&
 \left.+\frac{D_{12}}{M^2}\left(-\frac{2m^4}{3M^4}+\frac{50m^2}{3M^2}+\frac{4}{3}\right)+\frac{R}{M^2}\left(-2m^4-\frac{8M^4}{3}-\frac{14m^6}{3M^2}-\frac{10m^2M^2}{3}\right)\right]\\
I_{45}&=&\left[\frac{D_{11}}{M^3}\left(-\frac{8m^3}{M^2}-\frac{8m}{3}\right)+\frac{D_{22}}{M^3}\left(-\frac{8m^3}{M^2}+\frac{16m}{3}\right)+\frac{D_{12}}{M^3}\left(\frac{16m^3}{M^2}-\frac{8m}{3}\right)\right.\\&&
\left.\frac{D_1}{M^2}\left(\frac{8m^3}{M^2}-\frac{4m}{3}\right)+\frac{D_2}{M}\left(-\frac{8m^3}{M^2}+\frac{4m}{3}\right)\right]
\label{eq:I20}
\end{eqnarray*}
\vspace{-1.cm}
\begin{equation}
\phantom{1=1}
\end{equation}

Here $R,D_1,D_2,D_{11},D_{12},D_{22}$ are the analytic results of
contour integrations over the off-shell parameter $d\sigma$ in the
complex $\sigma$-plane. The results of these integrals are given as:
\begin{eqnarray}
\nonumber D_1&=&\int \frac{Md\sigma}{\Delta_1^2\Delta_2}\Delta_1=2\pi i\;\frac{1}{D(\widehat{q})}\\
\nonumber D_2&=&\int \frac{Md\sigma}{\Delta_1^2\Delta_2}\Delta_2=2\pi i\;\frac{2}{(2\omega)^3}\\
\nonumber D_{12}&=&\int \frac{Md\sigma}{\Delta_1^2\Delta_2}\Delta_1\Delta_2=2\pi i\;\frac{1}{2\omega}\\
\nonumber D_{11}&=&\int \frac{Md\sigma}{\Delta_1^2\Delta_2}\Delta_1^2=2\pi i\;\frac{1}{2\omega}\\
\nonumber D_{22}&=&\int \frac{Md\sigma}{\Delta_1^2\Delta_2}\Delta_2^2=2\pi i\;\frac{\displaystyle\omega^2-\frac{M^2}{2}}{\omega^3}\\
R&=&\int \frac{Md\sigma}{\Delta_1^2\Delta_2}=2\pi i\;\frac{M^2-12\omega^2}{4\omega^3(M^2-4\omega^2)^2}
\label{eq:theSi}
\end{eqnarray}
Final result for the BS normalizer has the form,
\begin{eqnarray}
N_V^{-2} &=&\frac{\pi^{5/2}}{72M^7\beta^3}e^{m^2/(2\beta^2)}\left\{
[G_{13}(m,M,A) + G_{11}(m,M,A)\beta^2 +
G_9(m,M,A)\beta^4]K_0\left(\frac{m^2}{2\beta^2}\right)\right.\nonumber\\
&&+[G_{13}(m,M,A) + G_{11}(m,M,A)\beta^2 +
G_9(m,M,A)\beta^4 + G_7(m,M,A)\beta^6]K_1\left(\frac{m^2}{2\beta^2}\right)\nonumber\\
&&+\left[H_5(m,M)\beta^6U\left(\frac{1}{2},-3,\frac{m^2}{\beta^2}\right)\right.\nonumber\\
&&+H_7(m,M)\beta^4U\left(\frac{1}{2},-2,\frac{m^2}{\beta^2}\right)\nonumber\\
&&\left.\left.+H_{11}(m,M)U\left(\frac{1}{2},0,\frac{m^2}{\beta^2}\right)\right]\beta^2A_3A_5\right\}
\label{eq:formNVm2}.
\end{eqnarray}
Here, $K_n(x)$ is the second class modified Bessel function, $U(a,b,x)$
is the confluent hypergeometric function, $G_n$ and $H_n$ are
polynomials of $n$-th degree in $m$ and $M$, and the $G_n$ are quadratic
functions of the $A_i$ coefficients.

In these expressions, $\phi(\hat{q})$ is a decaying function of
$\hat{q}^2$. Thus despite the fact that the integrands contain growing
factors like $\hat{q}^2$, the overall integrals converge, and can be
analytically integrated. Then, the $f_V^{i}$ $(i=0,...,5)$ can be
expressed in the following analytic form,
\begin{eqnarray}
\nonumber f_V^0&=&\frac{A_0}{M}N_V16\sqrt{\frac{2}{3}}\pi^{3/4}\beta^{3/2}(2m^2+3\beta^2)\\
\nonumber f_V^1&=&-A_1N_V8\sqrt{6}\pi^{3/4}\beta^{3/2}m\\
\nonumber f_V^2&=&0\\
\nonumber f_V^3&=&\frac{A_3}{M^2}N_V\frac{2\pi^{1/4}}{\sqrt{3}}\sqrt{\beta}
\left\{3M\sqrt{2\pi}\beta(-4m^2+M^2-28\beta^2)\phantom{\frac{2}{4}}\right.\\&&
\nonumber \left.+2e^{m^2/(4\beta^2)}m^2\left[2m^2K_0\left(\frac{m^2}{4\beta^2}\right)
\ +(2m^2-M^2+16\beta^2)K_1\left(\frac{m^2}{4\beta^2}\right)
\right]\right\}\\
\nonumber f_V^4&=&-\frac{A_4}{M^3}N_V\frac{4\pi^{1/4}}{\sqrt{3}}\sqrt{\beta}
\left\{m^2M^2
e^{m^2/(4\beta^2)}K_1\left(\frac{m^2}{4\beta^2}\right)
+2\sqrt{\pi}\beta^3\left[3\sqrt{2}M-8\beta U\left(-\frac{3}{2},-2,\frac{m^2}{2\beta^2}\right)
\right]\right\}\\
f_V^5&=&0
\label{eq:formfv}
\end{eqnarray}
with $N_{V}$ given in the previous equation.


\section{Results}
\label{sec:results}

\subsection{Numerical Calculation}
\label{subsec:numerical}

Eq.~(3.18) which expresses decay constants $f_V$ of vector
mesons in terms of the parameters $A\equiv($$A_0$, $A_1$, $A_2$, $A_3$,
$A_4$, $A_5)$ is a non linear function of the $A_i$'s. Then numerical
methods must be applied to solve the problem of determining the best
values of the $A_i$'s.

We used a simple `Mathematica' procedure for searching for accurate
values of the $A_i$ ($i=0,...5$). We defined the following auxiliary
function $W(A)$, which is positive definite, as,
\begin{equation}
W(A)=\sum\limits_V\left[f_V(A)-f_V^{EXP}\right]^2.
\label{eq:3.14}
\end{equation}
The summation in the above equation runs over the five ground state
vector mesons $\rho$, $\omega$, $\phi$, $\psi$ and $Y$ studied in this
work. $f_V^{EXP}$ are the central values of experimental data on decay
constants used (indicated in Table 2) which are calculable from the data
on the partial widths, $\Gamma(e^++e^-)$ for the five studied mesons in
~\cite{beringen12}. Their results are the following:
$\rho$(770): 7.04$\pm$0.06 keV,
$\omega$(782): 0.60$\pm$0.02 keV,
$\phi$(1020): 1.27$\pm$0.04 keV,
$\psi$(1S): 5.5$\pm$0.1 keV,
$Y$(1S): 1.34$\pm$0.02 keV.
They are related to our decay constants $f_V$ by formula,
\begin{equation}
\Gamma= \frac{4\pi\alpha^{2}e_Q^{2}|f_{V}|^{2}}{3M}.
\label{eq:width2decay}
\end{equation}
Here, $M$ is the meson mass, $\alpha$ is the QED coupling constant ({\em
i.e.} the fine structure constant), $e_Q$ plays the role of effective
electric charge of the meson and has values as listed after
Eq.(3.2)for different vector mesons.

From formula (\ref{eq:width2decay}) were obtained the data cited in
Table 1. We see that the error bars of experimental data on decay widths
given by PDG tables for the $\rho$, $\omega$, $\phi$, $\psi$ and $Y$
mesons represent, 0.8\%, 3.3\%, 3.1\%, 1.8\%, 1.5\%, respectively. We
can say that average error bars are 2.1\% or 0.05 keV. The average error
in decay constants derived from experimental data of the five mesons
studied is 1\%. As a general rule, while we are averaging data coming
from different measurements, we should give larger weight to more
precise data and lesser weight to less precise data. For instance,
$\rho$'s experimental error bar of $f_V$ is 0.4\%, while for $\omega$
meson error bar of $f_V$ is 1.6\%; {\em i.e.\/} $\rho$ has a higher
precision than $\omega$ meson. However, by using the $W(A)$ function, we
have fit the data for all mesons from $\rho$ to $Y$ at the same level.
This means that we can not expect a fitting with an average error
significantly less than $\sim1.6$\%.

From the numerical point of view, the problem reduces to finding values
of $A_i$'s such that $W(A)$ has a minimum, and that at such a minimum it
takes the value zero. We used the Mathematica package which has some
useful functions for minimizing. It is clear that the 6-dimensional
hypersurface $W(A)$ has many minima, but the only acceptable minima are those
minima for which $W(A)$ is very close to zero. The experimental values of
the $f_V$'s are 0.220, 0.195, 0.228, 0.410 and 0.708 GeV. Then, we adopt
as criterium of ``sufficient closeness'' to zero the value of
$W(A)$ which is less than $0.02^2=0.0004$ GeV$^2$.

Besides, it is important to ensure a degree of robustness of the
solution. It means, that if we change each of the experimental values
$f_V^{EXP}$ according to their error bars, then the values of the $A_i$
which minimize $W(A)$ remain near the corresponding values which
minimize at the central values $f_V^{EXP}$. This is the main criterion
of stability which must be satisfied for the model to be physically
acceptable. In other words, when point $(A)$ is within a box determined
by the averages and the error bars of the $A_i$, the function $W(A)$ has
a minimum very near to zero for $f_V$ in a box determined by the
experimental data of the $f_V$ and their error bars.

An additional check is done by evaluating the percent average of the
absolute values of the differences between the predicted $f_V$ values
from the experimental value $f_V^{EXP}$.

Using this method, we found that the values of coefficients should
respectively be: $A_0=1$, $A_1=0.006773$, $A_2=1.24011$,
$A_3=-0.414747$, $A_4=0.013611$, $A_5=-1.84191$ to predict the decay
constant values, $f_\rho=0.207440$ GeV, $f_\omega=0.206914$ GeV,
$f_\phi=0.230219$ GeV, $f_\psi=0.415707$ GeV, and $f_Y=0.758994$ GeV.
These decay constant values have an average error with respect to the
experimental data of 4\%.

The robustness of the model can be quantified in the following way. The
set of $A_i$ which, when replaced in $f_V(A)$ give a value ``near'' the
experimental value of the decay constant for each of the studied vector
mesons. We found that point $(A)$ is located within certain ``box'' of
sides $2\Delta A_i=0.008$. The center of the box is at the values given
in the last paragraph. All those results were obtained by randomly
choosing 22 sets of the $f_V$ within their experimental error bars and
finding in each case the point $A$ where $W(A)$ has a minimum near to
zero. Fig. 2 represents the $f_V$ amplitudes for all five studied
mesons, sets of dots were obtained by varying the $A_i$ coefficients
around their average values, procedure was done to show stability of
results.

The normalization factors $N_V$ were found to be:
$N_\rho=0.18707795$ GeV$^{-3}$,
$N_\omega=0.18624605$ GeV$^{-3}$,
$N_\phi=0.13413963$ GeV$^{-3}$,
$N_\psi=0.03270152$ GeV$^{-3}$, and
$N_Y=0.00495780$ GeV$^{-3}$.
Values of $f_V$ along with the contributions from various covariants and
experimental results are listed in Table 2. Comparison of our results
with those of other models and data is presented in Table 3.

\subsection{The results}
\label{subsec:results}

Formulas found in section \ref{sec:theory} express decay constants $f_V$
of vector mesons in terms of the constant parameters $A_0$, $A_1$,
$A_2$, $A_3$, $A_4$, $A_5$. Our model should be capable of predicting
the values of those parameters if one uses the known experimental values
of the decay constants for the $\rho$, $\omega$, $\phi$, $\psi$, $\psi'$
and $Y$ mesons. Expression for $f_V$ in Eq.~(3.12) and Eq.~(3.18) is a
linear function of the $A_i$'s (for i=0,1,2,3,4,5). However $f_{V}$
involves the BS normalizer $N_V$, which is evaluated by integrating with
respect to $\hat{q}$, is a highly non-linear function of the $A_i$'s.
Analytical expressions for the $A_i$'s as functions of quark masses and
other parameters corresponding to each of the vector mesons can not be
obtained. However, numerical methods give acceptable solutions of the
problem.

Expression for normalizer (\ref{eq:formNVm2}) has the form,
\begin{equation}
\frac{1}{N_V^2} = \sum\limits_{i=0}^5\sum\limits_{j=i}^5I_{ij}A_iA_j,
\label{eq:quadraticNv}
\end{equation}
where matrix elements $I_{ij}$ were given in Eq.~(\ref{eq:I20}).

Contributions $f_V^i$ to the decay constants, given by
Eq.~($\ref{eq:formfv}$), by definition are proportional to $A_i$,
\begin{equation}
f_V = \sum\limits_{i=0}^5f_V^i \equiv \sum\limits_{i=0}^5\frac{f_{V_i}}{N_V}A_i.
\end{equation}

Our idea is that the $A_i$ can be obtained by fitting formulas to
experimental results,
\begin{equation}
\frac{\sum\limits_{i=0}^5f_{V_i}A_i}{\sqrt{\sum\limits_{i=0}^5
\sum\limits_{j=i}^5I_{ij}A_iA_j}} = f_V^{EXP}.
\label{eq:equationAi}
\end{equation}

(see Eq.~(\ref{eq:Nvm2}) and Eq.~(\ref{eq:I20})). We must notice that
\ref{eq:equationAi} is a homogenous function of the $A_i$, fact that
precludes finding solutions for the $A_0,... A_5$ by using available
data of the five considered mesons. However, a nonhomogenous system of
equations can be constructed by putting $A_0=1$ and leaving as unknowns
$A_1,... A_5$. With this procedure it is sufficient to consider the five
mesons $\rho$, $\omega$, $\phi$, $\psi$, and $Y$. System of five
equations is obtained by introducing in Eq.~(\ref{eq:equationAi}) the
appropriate parameters of the five mesons.

For the $f_V^{EXP}$ we used the results from PDG tables
\cite{beringen12}. We see that the ``error bars'' on data for $f_{V}$
for $\rho,\omega,\phi,\psi$ and $Y$ mesons are 0.4\%, 1.7\%, 1.6\%,
0.9\% and 0.8\% respectively, whose average is about 1.1\%.

System of five equations has many solutions, several of them complex.
Complex solutions appear in conjugate pairs. Available algorithms allow
finding all solutions with very high precision. In this way the percent
average of the absolute values of the difference between the calculated
$f_V$ and the $f_V^{EXP.}$ experimentally found, is much lower than the
error bars of experiments. However the values of the $A_i$ found were
complex, which lead to complex $f_V$'s. We made different checks and
selected the parameter set giving the results shown in Table 2 which
predicts the experimental values of all the $f_V$'s with less precision.
The $f_V$'s predicted for the considered mesons match approximately with
the experimental values. Regarding meson $\psi'$, our theoretical value
is 0.1798 GeV, while value calculated from experiments is 0.2200 GeV,
with error 18\%.


\begin{table*}[tbp]
\caption{\label{tab:i} Some parameters of vector mesons. $M$ is meson
mass, $m$ is mass of quark constituent, $\beta$ is the inverse range
parameter appearing in BS wave function, $\Gamma^{EXP.}$ is the
experimental value of width~\cite{beringen12} and $f_V^{EXP.}$ is the
decay constant deduced from $\Gamma^{EXP.}$ by using formula
Eq.~(\ref{eq:width2decay}). $e_Q$ arises from the flavor configuration
of individual vector mesons.}
\label{tab:mesonparameters}
\begin{center}
\begin{tabular}{|l|cccccc|}
\hline
Meson         & $m$ (GeV) & $M$ (GeV)           & $e_Q$         & $\beta$  & $\Gamma^{EXP.}$ (keV) & $f_V^{EXP.}$ (GeV) \\
\hline
$\rho(770)$   & 0.265     & 0.7755$\pm$0.0003   & $1/\sqrt{2}$  & 0.265294 & 7.04$\pm$0.06  & 0.2201$\pm0.0009$ \\
$\omega(782)$ & 0.265     & 0.7827$\pm$0.0001   & $1/\sqrt{18}$ & 0.2661   & 0.60$\pm$0.02  & 0.195$\pm0.003$   \\
$\phi(1020)$  & 0.415     & 1.01946$\pm$0.00002 & $1/3$         & 0.293347 & 1.27$\pm$0.04  & 0.228$\pm0.003$   \\
$J/\psi(1S)$  & 1.532     & 3.09692$\pm$0.00001 & $2/3$         & 0.444214 & 5.5$\pm$0.1    & 0.410$\pm0.003$   \\
$Y(1S)$       & 4.9       & 9.4603$\pm$0.0003   & $1/3$         & 0.721345 & 1.34$\pm$0.02  & 0.708$\pm0.005$   \\
\hline
\end{tabular}
\end{center}
\end{table*}
\vspace{1.5cm}
\begin{table*}[tbp]
\caption{\label{tab:ii} Decay constant $f_V^{TH.}$ values (in GeV) for $\rho$,
$\omega$, $\phi$, $\psi$ and $Y$ mesons in BSE-CIA with the individual
contributions $f_V^0$, $f_V^1$, $f_V^2$, $f_V^3$, $f_V^4$, $f_V^5$ from
various Dirac covariants along with the contributions from LO and NLO
covariants and also their \% contributions for parameter set: $A_0 = 1$,
$A_1 =0.00677$, $A_2 = 1.24011$, $A_3 = -0.41474$, $A_4 = 0.01361$, $A_5
= -1.84191$ (with average error with respect to the experimental data
of 4\%)}
\begin{center}
\begin{tabular}{|c|c|c|c|c|c|c|c|c|c|c|}
\hline
         & $f_V^0$ & $f_V^1$   & $f_V^3$ & $f_V^4$   & $f_V^{LO}$ & $f_V^{NLO}$ & $f_V^{LO}$(\%) & $f_V^{NLO}$(\%) & $\bm{f_V^{TH.}}$ & $\bm{f_V^{EXP}}$ \\ \hline
$\rho$   & 0.1156  & -0.00068  & 0.09336  & -0.00026    & 0.1149     & 0.0931      & 55\%           & 45\%            & \textbf{0.2080}  & \textbf{0.2200}  \\ \hline
$\omega$ & 0.1155  & -0.000689  & 0.093   & -0.000258 & 0.1148     & 0.0927      & 56\%         & 44\%          & \textbf{0.2075}  & \textbf{0.1952}  \\ \hline
$\phi$   & 0.1461  & -0.00104  & 0.086   & -0.000288 & 0.1450      & 0.0859      & 63\%           & 37\%            & \textbf{0.2302}  & \textbf{0.2285}  \\ \hline
$\psi$   & 0.352   & -0.00321  & 0.06257 & -0.000254 & 0.3487     & 0.06232     & 84.8\%         & 15.2\%          & \textbf{0.411}  & \textbf{0.4104}  \\ \hline
$Y$      & 0.6617  & -0.00628  & 0.05268 & -0.000224 & 0.6555     & 0.0525      & 92.6\%         &  7.4\%          & \textbf{0.7079}  & \textbf{0.7080}  \\ \hline
\end{tabular}
\end{center}
\end{table*}
\vspace{1.5cm}
\begin{table*}[tbp]
\caption{\label{tab:iii} Calculated decay widths $\Gamma^{TH.}$ for the
process, $V\rightarrow\gamma*\rightarrow e^{+}+ e^{-})$ (in keV) for
$\rho$, $\omega$, $\phi$, $\psi$ and $Y$ mesons in BSE-CIA along with
their experimental values \cite{beringen12}}
\begin{center}
\begin{tabular}{|c|c|c|}
\hline
                      & $\Gamma^{TH.}$ (keV) & $\Gamma^{EXP.}$ (keV) \\ \hline
$\rho(770)$           & 8.952                & 7.04$\pm$0.06 \\ \hline
$\omega(782)$         & 0.642                & 0.60$\pm$0.02 \\ \hline
$\phi(1020)$          & 1.294                & 1.27$\pm$0.04 \\ \hline
$\psi(1S)(3096)$      & 5.414                & 5.5$\pm$0.1 \\ \hline
$Y(1S)(9460)$         & 1.345                & 1.34$\pm$0.02 \\ \hline
\end{tabular}
\end{center}
\end{table*}
\vspace{1.5cm}

\begin{table*}[tbp]
\caption{\label{tab:iv} Decay constant $f_V$ values (in GeV) for $\rho$,
$\omega$, $\phi$, $\psi$ and $Y$ mesons in BSE-CIA and their comparison
with other models and data)}
\begin{center}
\begin{tabular}{|c|c|c|c|c|c|}
\hline
           & $f_\rho$         & $f_\omega$   & $f_\phi$     & $f_\psi$ & $f_Y$ \\ \hline
$BSE-CIA$  & 0.208            & 0.207        & 0.230        &  0.411   & 0.7079   \\ \hline
$BSE [4]$  & 0.215            &              & 0.224        &          &        \\ \hline
$SDE [2]$  & 0.163            &              &              &          &        \\ \hline
$SDE [7]$  & 0.207            &              & 0.259        &          &        \\ \hline
$BSE [8]$  &                  &              &              & 0.459    & 0.498   \\ \hline
$Exp.[22]$ & 0.2201$\pm$0.0009& 0.195$\pm$0.003 & 0.228$\pm$0.003 & 0.410$\pm$0.003& 0.708$\pm$0.005 \\ \hline
\end{tabular}
\end{center}
\end{table*}
\vspace{1.5cm}


\begin{figure*}[bt]
\begin{center}
\includegraphics[height=0.5\linewidth,width=0.8\linewidth,angle=0]{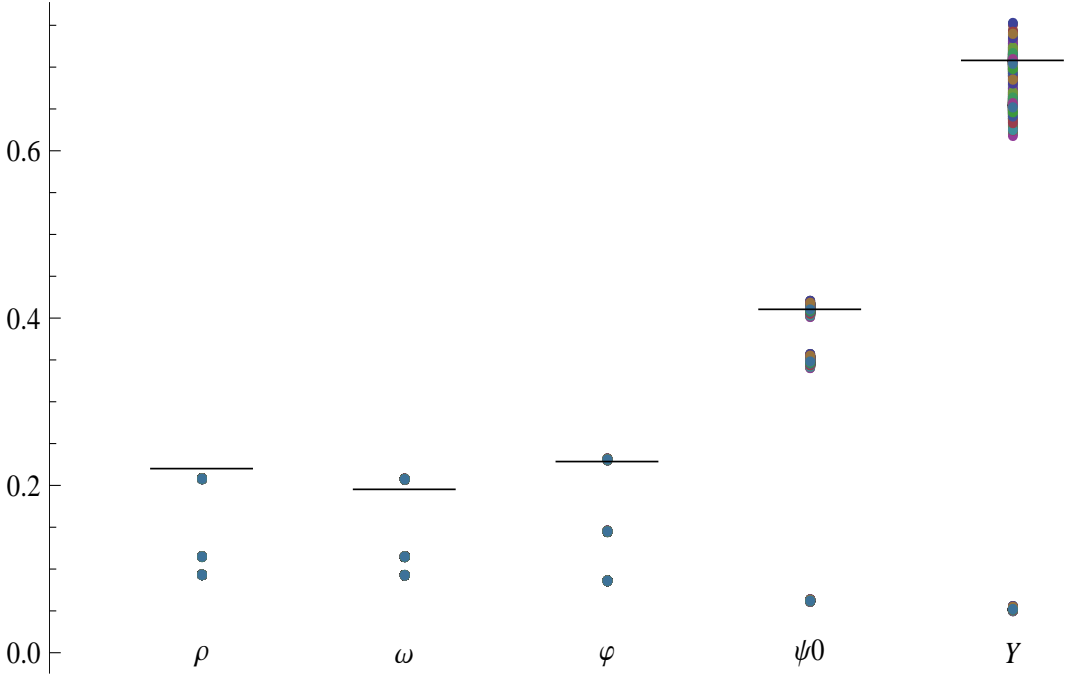}
\caption{Amplitudes $f_V$ for all studied mesons are represented.
Vertical scale is in units of Gev. Short horizontal lines are values
obtained from experimental data. Lower sets of points are the NLO
contributions obtained from our model. Intermediate sets of points are
our LO results. Upper sets of points are our theoretical values of
$f_V=f_V^{LO}+f_V^{NLO}$. It is concluded that same set of $A_i$
coefficients predicts simultaneously the $f_V$ for all five studied
mesons and that uncertainties in the $A_i$ are strongly ``amplified''
for $Y$ and $\psi_0$.}
\end{center}
\end{figure*}

\begin{figure*}[bt]
\begin{center}
\includegraphics[height=0.4\linewidth,width=0.5\linewidth,angle=0]{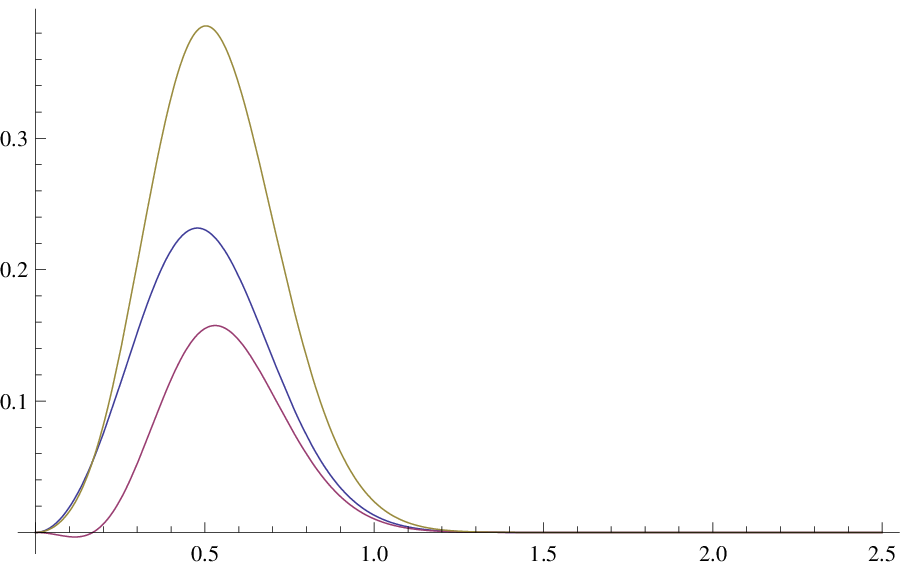}\includegraphics[height=0.4\linewidth,width=0.5\linewidth,angle=0]{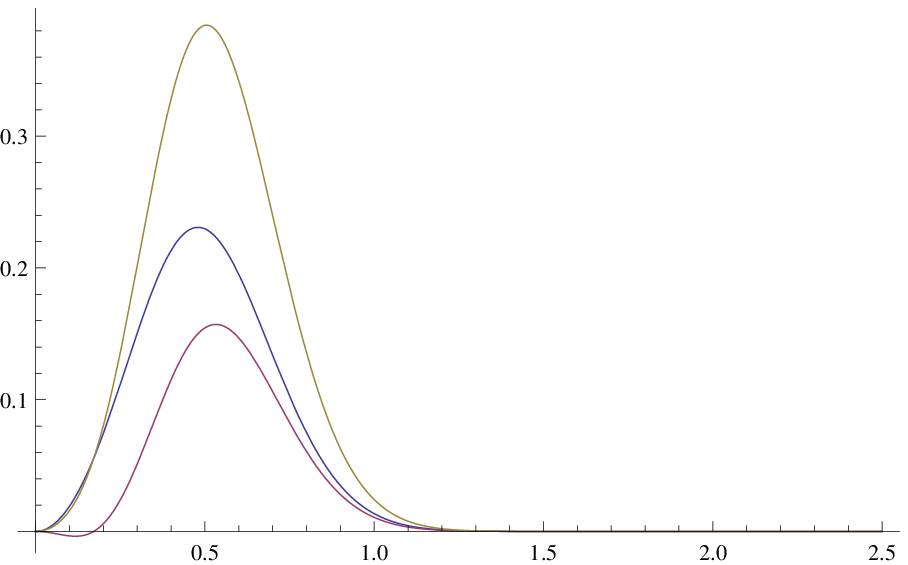}
\includegraphics[height=0.4\linewidth,width=0.5\linewidth,angle=0]{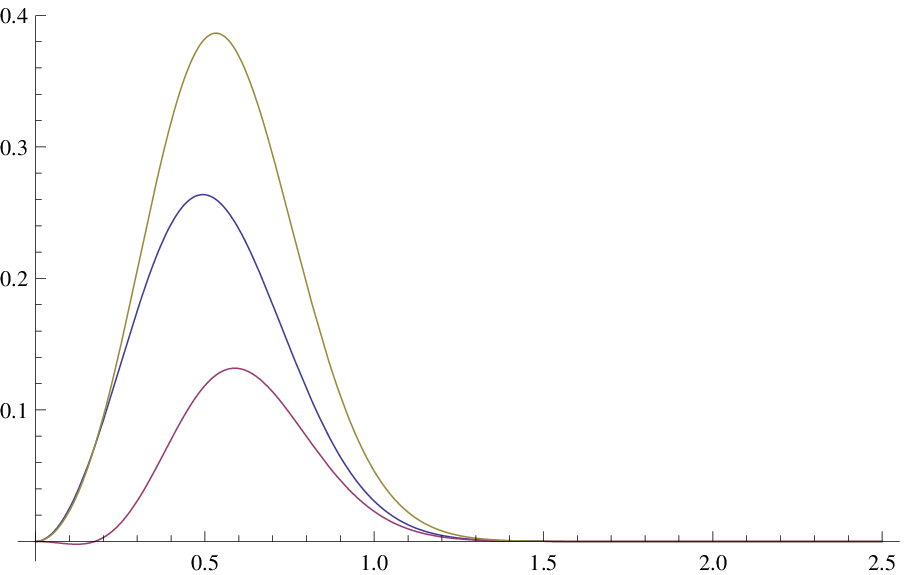}\includegraphics[height=0.4\linewidth,width=0.5\linewidth,angle=0]{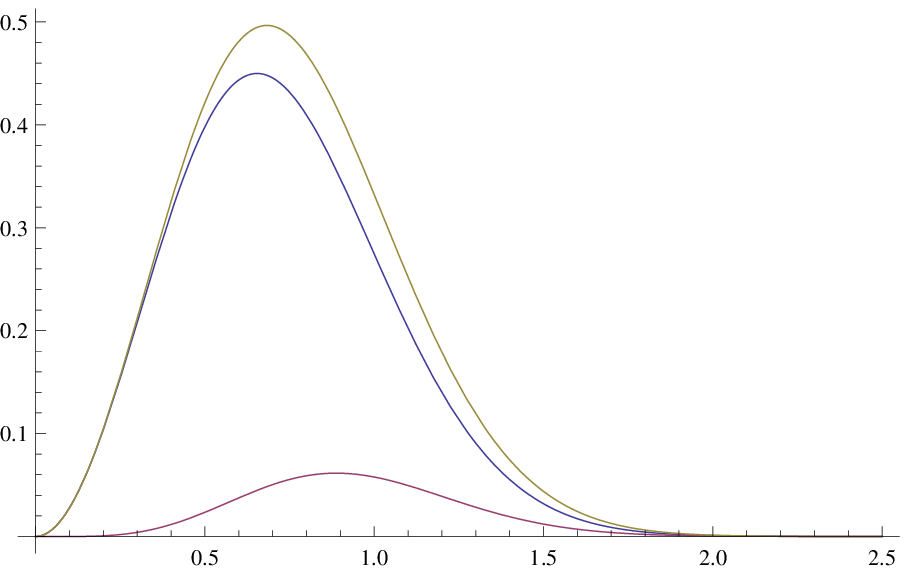}
\includegraphics[height=0.4\linewidth,width=0.5\linewidth,angle=0]{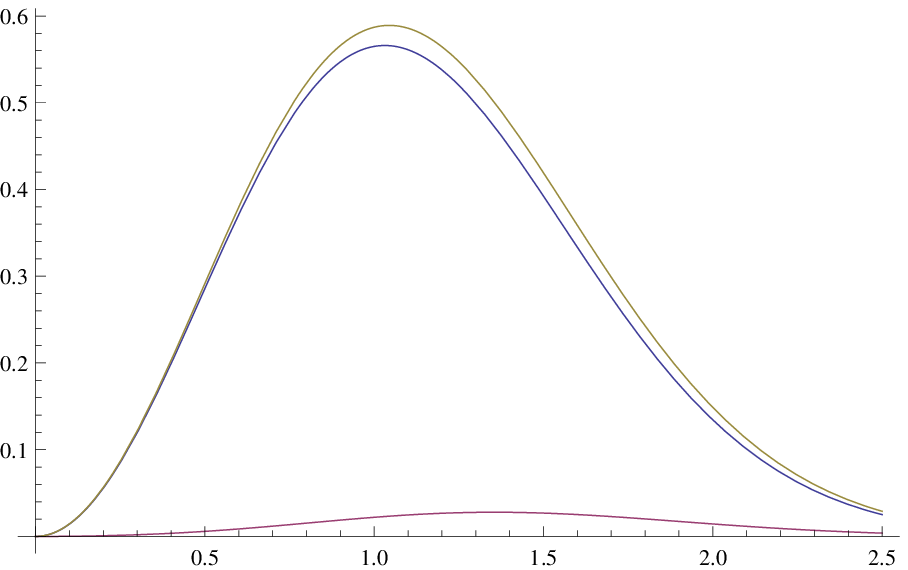}
\caption{Plots of integrand functions which give rise to $f_V^{LO}$,
$f_V^{NLO}$, and $f_V$, as functions of $\widehat{q}$, for the five
vector mesons studied, respectively.}
\end{center}
\end{figure*}


\section{Discussion}
\label{sec:disc}

In this paper we have calculated the decay constants $f_V$ of vector
mesons $\rho$, $\omega$, $\phi$, $\psi$ and $Y$ in BSE under Covariant
Instantaneous Ansatz (CIA) using Hadron-quark vertex function which
incorporates various Dirac covariants order-by-order in powers of
inverse of meson mass within its structure in accordance with a recently
proposed power counting rule from their complete set. This power
counting rule suggests that the maximum contribution to any meson
observable should come from Dirac structures associated with Leading
Order (LO) terms alone, followed by Dirac structures associated with
Next-to-Leading Order (NLO) terms in the vertex function. Incorporation
of all these covariants is found to bring calculated $f_V$ values much
closer to results of experimental data \cite{beringen12} and some recent
calculations
\cite{alkofer02,ivanov99,close02,li08,cvetic04,hwang97,alkofer01,wang08}
for $\rho$, $\omega$, $\phi$, $\psi$ and $Y$ mesons. The $f_V$ predicted
are within the error bars of experimental data for each one of these
five mesons.

The results for $\rho$, $\omega$, $\phi$, $\psi$ and $Y$ mesons with
parameter set:
$A_0 = 1$,
$A_1 = 0.007\pm0.001$,
$A_2 = 1.240\pm0.001$,
$A_3 = -0.415\pm0.001$,
$A_4 = 0.014\pm0.001$,
$A_5 = -1.842\pm0.001$
(giving $f_V$ values with average error with respect to experimental
data of 4\%) are presented in Table 2. Comparison with experimental data
and other models is shown in Table 2.

In Fig. 3 we are plotting vs $\hat{q}$ the integrands of $f^{LO}_V$,
$f^{NLO}_V$ and $f_V$ for each of studied mesons. Those plots show that
the contribution to $f_V$ from NLO covariants is smaller than the
contribution from LO covariants for $\rho$, $\omega$, $\phi$, $\psi$ and
$Y$ mesons. And for $\psi$ and $Y$ mesons, NLO contribution is
negligible in comparison to LO contribution. Then, it is concluded from
Table 2 that as far as the various contributions to decay constants
$f_V$ are concerned, for $\rho$ and $\omega$ mesons, the LO terms
contribute only ~55\%, while NLO terms ~45\%. However as one goes to
$\phi$ meson, the LO contribution increases to ~62.8\%, while NLO
contribution is 37.2\%. But as one goes to heavy ($c\overline{c}$ and
$b\overline{b}$) mesons, for $\psi$ meson, LO contribution is 84.8\%,
while NLO contribution is 15.2\%, and for $Y$ meson the LO contribution
is 92.6\%, while NLO contribution reduces to just 7.4\%. Thus the drop
in contribution to decay constants from NLO covariants vis-a-vis LO
covariants is more pronounced for heavy mesons $\psi$ and $Y$. And among
the two LO covariants, it can be seen that the most leading covariant
$i\gamma_{\mu}$ contributes the maximum for all vector mesons from
$\rho$ to $Y$. These results on decay constants $f_{V}$ for vector
mesons are completely in conformity with the corresponding results on
decay constants $f_{P}$ for pseudoscalar mesons, $K,D,D_{S}$ and $B$
done recently\cite{bhatnagar11} where it was also noticed that the NLO
contribution is much smaller than the LO contribution for hevier mesons
like $D,D_{S}$ and $B$, where the contribution drops from 10\% to 4\%,
and the most leading covariant was found to be $\gamma_{5}$.

This is in conformity with the power counting rule according to which 
the leading order covariants, $\gamma_5$ and $i\gamma_5(\gamma\cdot 
P)(1/M)$ (associated with coefficients $A_0$ and $A_1$) should 
contribute maximum to decay constants followed by the next-to-leading 
order covariants, $-i\gamma_5(\gamma\cdot q)(1/M)$ and 
$-\gamma_5[(\gamma\cdot P)(\gamma\cdot q)-(\gamma\cdot q)(\gamma\cdot 
P)](1/M^2)$ (associated with coefficients $A_2 $ and $A_3$) in the BS 
wave function.

We observe in Fig. 2 that though LO and NLO Dirac covariants are
sufficient to correctly predict amplitudes for $\phi$ and $\psi_0$
vector mesons, but only LO covariants are sufficient for $Y$ meson. But
for $\rho$ and $\omega$ mesons, the LO and NLO Dirac covariants are not
sufficient to predict accurately their amplitudes and is thus necessary
to include even higher order NNLO Dirac covariants in their hadron-quark
vertex functions. This can also be seen from Fig. 3.

However, the numerical results for $f_V$ for equal mass vector mesons,
obtained in our framework with use of leading order Dirac covariants
along with the next to leading order Dirac covariants, along with a
similar calculation for $f_{P}$ done recently \cite{bhatnagar11} for
pseudoscalar mesons demonstrates the validity of our power counting
rule, which also provides a practical means of incorporating various
Dirac covariants in the BS wave function of a hadron. By this rule, we
also get to understand the relative importance of various covariants to
calculation of meson observables. This would in turn help in obtaining a
better understanding of the hadron structure.


\section*{Acknowledgments}

One of the authors (SB) is thankful to Prof. S-Y. Li, Shandong University,
China for discussions. JM thanks the support from Programa de
Sostenibilidad University of Antioquia.


\end{document}